\pdfoutput=1 

\documentclass[aps,pra,twocolumn,groupedaddress,showpacs,amsmath]{revtex4}

\usepackage{graphicx}
\usepackage{amssymb}
\usepackage{comment}

\usepackage{bm} 

\newcommand*{\br}{\mathbf{r}}

\newcommand*{\bk}{\mathbf{k}}
\newcommand*{\bq}{\mathbf{q}}
\newcommand*{\bv}{\mathbf{v}}

\newcommand*{\cL}{{\cal L}}
\newcommand*{\cH}{{\cal H}}

\newcommand*{\Phop}{\Phi^{\vphantom{\dagger}}}
\newcommand*{\Phdop}{\Phi^\dagger}

\newcommand{\vect}[1]{\mathbf{#1}}

\begin{document}

\title{Long wavelength spin dynamics of ferromagnetic condensates}

\author{Austen Lamacraft}
\affiliation{Department of Physics, University of Virginia,
Charlottesville, VA 22904-4714 USA}
\date{\today}
\email{austen@virginia.edu}

\begin{abstract}

We obtain the equations of motion for a ferromagnetic Bose condensate of arbitrary spin in the long wavelength limit. We find that the magnetization of the condensate is described by a non-trivial modification of the Landau-Lifshitz equation, in which the magnetization is advected by the superfluid velocity. This hydrodynamic description, valid when the condensate wavefunction varies on scales much longer than either the density or spin healing lengths, is physically more transparent than the corresponding time-dependent Gross-Pitaevskii equation. We discuss the conservation laws of the theory and its application to the analysis of the stability of magnetic helices and Larmor precession. Precessional instabilities in particular  provide a novel physical signature of dipolar forces. Finally, we discuss the anisotropic spin wave instability observed in the recent experiment of Vengalattore \emph{et. al.} (Phys. Rev. Lett. {\bf 100}, 170403, (2008)).



\end{abstract}


\maketitle

\section{Introduction}

Perhaps the most dramatic way in which Bose condensation in the alkali gases differs from its counterpart in $^4$He (or, for that matter, in conventional $s$-wave superconductors) is that the condensed particles have non-zero spin~\cite{ketterle2000}. Magnetic trapping results in a gas in which the spin state of the atoms can be described using the adiabatic approximation, with the atoms remaining in a particular hyperfine level relative to the local magnetic field as they move around the trap. Although there are several interesting consequences of the non-uniformity of the field~\cite{ho1996}, far richer behavior results when atoms are optically trapped, allowing the full consequences of rotational invariance to be realized. The experimental preparation of a long-lived gas in a particular hyperfine multiplet may be more difficult for some atoms than others, but the theorist is nevertheless called upon to answer the question: \emph{what are the properties of the higher spin Bose condensates, and how will they manifest themselves in the ultracold laboratory?}

Starting with Refs.~\cite{ohmi1998,ho1998,koashi2000,ciobanu2000}, a number of investigations have explored the possible magnetic phases of these condensates. The dynamics of such an ordered phase is generally described by a non-linear equation for the motion of the order parameter on some manifold of symmetry-broken states, but this line of thought  has not been much pursued.  This is partly due to the existence of a dynamical description of Bose condensates valid in the dilute limit, namely the time-dependent Gross-Pitaevskii (GP) equation, which may be straightforwardly extended to the multicomponent case. A hydrodynamical description of the `slow' degrees of freedom -- generally the order parameter and any conserved quantities -- is nevertheless desirable both for physical transparency and simplicity, at the modest cost of eliminating certain uninteresting high frequency motions. The goal of this work is to develop and apply such a description for the case of a ferromagnetic condensate. 

The structure of the remainder of the paper is as follows. In the next section we derive the equations of motion valid in the long wavelength limit starting from the Gross-Pitaevskii Lagrangian for a ferromagnetic condensate. After discussing the conservation laws of the theory the equations of motion are used to study the stability of magnetic helices (Section~\ref{sec:helix}), Larmor precession in the presence of dipolar forces (Section~\ref{sec:larmor}) and spin waves in the presence of Larmor averaged dipolar forces (Section~\ref{sec:ave}). The relevance of the last calculation to the experiment of Ref.~\cite{Vengalattore2008} is briefly discussed.

\section{Equations of motion}

Our starting point is the Gross-Pitaevskii Lagrangian density ($\hbar=m=1$)
\begin{eqnarray}\label{GP_L}
\cL&=&i\Phdop\partial_t\Phop-\cH(\Phdop,\Phop)\nonumber\\
\cH&=&\frac{1}{2}\left[\nabla\Phdop\nabla\Phop+c_0\left(\Phdop\Phop\right)^2+c_2\left(\Phdop \vect{S\vect{}}\Phop\right)^2\right],
\end{eqnarray}
where $\Phop$ is a $2s+1$ component spinor and $\vect{S}$ the spin-$s$ matrices. The quartic terms are appropriate to the description of interatomic interactions in a spin-1 condensate such as $^{87}$Rb, where $c_2<0$ favors ferromagnetism~\cite{ohmi1998,ho1998}. The description of higher spin condensates requires more parameters but the phase diagram always includes a ferromagnetic phase~\cite{koashi2000,ciobanu2000}.

We are going to work in the low energy limit where both interaction terms are fully satisfied. This is appropriate to the limit where the condensate wavefunction varies on scales much longer than either the density or spin healing length, or equivalently, the superfluid velocity is small compared to the speed of propagation of sound or spin waves. This is analogous to the incompressible limit used to describe normal fluids at low local Mach number. The density interactions demand $\rho(\br)=\Phdop\Phop=\mathrm{const.}$, which we set equal to unity from now on, while the spin interactions, assumed ferromagnetic, demand that the polarization be maximal at each point, so that the spinor $\Phop$ is a spin coherent state. Assuming that the interactions dominate all other terms in the Hamiltonian allows other terms describing e.g. Zeeman and dipole-dipole interactions to be included in a controlled way by simply evaluating them on the constrained ferromagnetic manifold. A similar approach to the static case was introduced in~\cite{kuratsuji2001,takahashi2007}. 

In this limit the equations of motion for the unit vector $\vect{n}(\vect{r},t)$ describing the local magnetization $s\vect{n}=\Phdop\vect{S}\Phop$ will be shown to be
\begin{subequations}
\begin{equation}\label{LLE}
\frac{D\vect{n}}{Dt}-\frac{1}{2}\vect{n}\times \nabla^2\vect{n}=0
\end{equation}
\begin{eqnarray}\label{v_fix}
\nabla\cdot \vect{v}=0,\qquad
\nabla\times\vect{v}=\frac{s}{2} \varepsilon_{\alpha\beta\gamma} n_\alpha \nabla n_\beta \times \nabla n_\gamma,
\end{eqnarray}
\end{subequations}
where $D/Dt$ denotes the usual Eulerian derivative $\partial_t+\vect{v}\cdot\nabla$. The first equation is a modified Landau-Lifshitz equation (LLE), that accounts for the advection of the magnetization by the superfluid velocity $\bv$. The other two equations determine this flow from the condition of vanishing divergence (incompressibility) and the \emph{Mermin-Ho relation}~\cite{mermin1976} that fixes the vorticity $\bm{\omega}\equiv\nabla\times\vect{v}$. The lines of vorticity coincide with the lines of constant $\vect{n}$. Eqs.~(\ref{v_fix}) fix $\vect{v}$ up to a some potential contribution $\nabla\psi$ and $\nabla^2\psi=0$. This should be chosen so that the normal component of the velocity vanishes at the boundary of the flow. Note that the dynamics of the system depends crucially on the spin $s$, with the usual LLE being recovered in the $s\to 0$ limit.


\emph{Variational principle} The dynamical equations may be found by a variational principle through an appropriate parametrization of the constraint manifold. We use the parametrization
\[\Phop=\Phop_{\vect{n}}e^{i\theta}\]
where $\Phop_{\vect{n}}$ is a normalized eigenstate of $\vect{n}\cdot\vect{S}$ with eigenvalue $s$. As expected there is some gauge freedom in how the overall phase of the state is apportioned between $\Phop_{\vect{n}}$ and the phase factor $e^{i\theta}$. For instance, the velocity is given by $\vect{v}=\nabla\theta-\vect{a}$, where $\vect{a}\equiv i\Phdop_{\vect{n}}\nabla\Phop_{\vect{n}}$. This vector potential depends on the gauge choice, though its curl does not
\[\nabla\times\vect{a}=i\nabla\Phdop\times\nabla\Phop=-\frac{s}{2} \varepsilon_{\alpha\beta\gamma} n_\alpha \nabla n_\beta \times \nabla n_\gamma\]
which is just the Memin-Ho relation.
Incompressibility translates to the constraint
\begin{equation}\label{incomp_con}
\nabla\cdot(\nabla\theta-\vect{a})=0,
\end{equation}
while the Hamiltonian and Lagrangian Eq.~(\ref{GP_L}) take the form
\begin{eqnarray*}
\cH&=&\left[\frac{1}{4}s\left(\nabla\vect{n}\right)^2+\frac{1}{2}\left(\nabla\theta-\vect{a}\right)^2\right]\nonumber\\
\cL&=&(a_t-\dot\theta)-\cH.
\end{eqnarray*}
We obtain the equations of motion by a variation of the associated action, bearing in mind Eq.~(\ref{incomp_con}). The non-trivial part is finding the variation of the terms involving $a_\mu$ ($\mu=t,\br$), which depends upon $\vect{n}$, without introducing a specific parametrization. This is accomplished by writing the field strength $\partial_\mu a_\nu-\partial_\nu a_\mu=-s\varepsilon_{\alpha\beta\gamma} n_\alpha \partial_\mu n_\beta  \partial_\nu n_\gamma$ including a fictitious extra coordinate $u$ with $\vect{n(t,\br,u)=}\vect{n}(t,\br)+u\delta\vect{n}(t,\br)$. Then 
\[\delta a_\mu=\partial_u a_\mu=s \vect{n}\times \partial_\mu\vect{n}\cdot\delta\vect{n}+i\partial_\mu \Phdop_\vect{n}\delta\Phop_\vect{n}\]
The variational derivative of the action with respect to $a_\mu$ is just the current $j_\mu\equiv\left(1,\vect{v}\right)$, so we have
\begin{eqnarray*}\label{aj_vary}
\int d\br dt\, \delta a_\mu j_\mu=\int d\vect{r}dt\left[sj_\mu \vect{n}\times \partial_\mu\vect{n}\cdot\delta\vect{n}-i\Phdop_\vect{n}\delta\Phop_\vect{n}\partial_\mu j_\mu\right],
\end{eqnarray*}
where integration by parts has been used to obtain the second term. The boundary term is equal to zero as $\delta\vect{n}$ vanishes there. Since the current is conserved by the incompressibility condition, the second term drops out and we have our variation. A final subtlety is that when we vary $\delta \vect{n}$ we also have to take into account the change $\delta\theta$ implied by Eq.~(\ref{incomp_con})
\[\nabla^2\delta\theta=\nabla\cdot\delta \vect{a}\]
This variation is handled in precisely the same way, but now the boundary term is zero due to the vanishing of the normal component of the velocity at the boundary.

Together with the variation of the first term of the Hamiltonian, the above variation readily gives the equation of motion Eq.~(\ref{LLE}).

\emph{Conservation laws}  Let us start by noting that Eq.~(\ref{LLE}) can be understood as the conservation equation for the magnetization $\partial_t\vect{n}+\partial_i\vect{J}_i
=0$ with the spin current
\[\vect{J}_i\equiv \vect{n}v_i-\frac{1}{2}\vect{n}\times\partial_i\vect{n}.\]
Another conservation law follows by first considering the equation of motion for the vorticity $\bm{\omega}\equiv\nabla\times\vect{v}$~\cite{papanicolaou1991}
\begin{eqnarray*}\label{vorticity_eom}
\frac{D\omega_i}{Dt}=-\frac{s}{2}\varepsilon_{ijk}\partial_k\partial_l\sigma_{lj}.\nonumber\\
\sigma_{ij}\equiv\frac{1}{2}\delta_{ij}\partial_k\vect{n}\cdot\partial_k\vect{n}-\partial_i\vect{n}\cdot\partial_j\vect{n}.
\end{eqnarray*}
%
%
which can be used to check that the \emph{hydrodynamic impulse} defined by
\begin{equation*}\label{impulse}
\vect{I}\equiv\frac{1}{2}\int \vect{r}\times \bm{\omega},
\end{equation*}
is a constant of the motion, as it is for both the LLE (where $\bm{\omega}$ is defined from the Mermin-Ho relation) and the Euler equation for incompressible flow. If external forces act on the fluid, they give the rate of change of the impulse. For the case of rigid walls that we have considered the total momentum $\int \vect{v}$ of the fluid naturally remains zero, with the reaction force of the walls being transmitted instantaneously  to the body of the fluid due to incompressibility~\cite{saffman1995}.

Finally we have the topological \emph{helicity} invariant
\begin{equation*}
\cH\equiv\frac{1}{16\pi^2}\int \vect{v}\cdot\bm{\omega}=-\frac{1}{16\pi^2}\int \varepsilon_{ijk}(\Phdop\partial_i\Phop)(\partial_j\Phdop\partial_k\Phop), 
\end{equation*}
which again is known in both hydrodynamics and magnetism~\cite{moffatt1969,dzyloshinskii1979}. For a condensate it is equal to the integer Brouwer degree of the map $g:\mathbb{R}^3\to SU(2)$ given by $\Phop(\br)=g(\br)\Phop_0$ for some 	fiducial state $\Phop_0$. Topological defects have been sought in the spin-1/2 case, but so far without success~\cite{Al-Khawaja2001,herbut2006}.



\section{Instability of a magnetic helix} \label{sec:helix}

We now apply these equations to the analysis of the stability of helical configurations of the magnetization
\begin{equation*}
\vect{n}_0(\xi)=\vect{e}_z\cos\theta+\sin\theta\left(\vect{e}_x\cos \xi+\vect{e}_y\sin \xi\right),
\end{equation*}
with $\xi\equiv qz-\omega_0t$. The instabilities of these configurations were studied in the recent experiment Ref.~\cite{Vengalattore2008}.  Since there is only a single wavevector present $\bm{\omega}=\vect{v}=0$. Substitution into Eq.~(\ref{LLE}) gives  $\omega_0= \frac{1}{2}q^2\cos\theta$. To analyze the stability of this configuration we introduce the orthonormal Frenet-Serret frame $\{\vect{n}_0,\vect{e}_1,\vect{e}_2\}$ consisting of axes parallel to $\vect{n}_0$, $\partial_\xi\vect{n}_0$, and the vector perpendicular to both. These satisfy
\begin{eqnarray*}
\partial_\xi\left(\begin{array}{c}\vect{n}_0 \\\vect{e}_1 \\\vect{e}_2\end{array}\right)&=&\left(\begin{array}{ccc}0 & \kappa & 0 \\-\kappa & 0 & \tau \\0 & -\tau & 0\end{array}\right)\left(\begin{array}{c}\vect{n}_0 \\\vect{e}_1 \\\vect{e}_2\end{array}\right)
\end{eqnarray*}
%
with curvature $\kappa=\sin\theta$ and torsion $\tau=\cos\theta$. Small deviations from the helix are then written as $\vect{n}=\vect{n}_0+\eta_1\vect{e}_1+\eta_2\vect{e}_2$. It is instructive to first discuss the prediction of the usual LLE without the advective term, which gives the coupled equations of motion
\begin{eqnarray*}
\dot\eta_1&=&-\frac{1}{2}\nabla^2\eta_2-q\cos\theta\partial_z\eta_1-\frac{q^2}{2}\sin^2\theta\eta_2\nonumber\\
\dot\eta_2&=&\frac{1}{2}\nabla^2\eta_1-q\cos\theta\partial_z\eta_2,
\end{eqnarray*}
with dispersion $\Omega(k)=q\cos\theta k_z\pm\sqrt{\frac{k^2}{2}\left[\frac{k^2}{2}-\frac{q^2}{2}\sin^2\theta\right]}$, revealing an instability for $0<k<q\sin\theta$. Note that the growth rate of the unstable modes is isotropic. The non-linear evolution of the helix in the one-dimensional integrable case is discussed in Ref.~\cite{ragan1998}, but let us now return to the full equation of motion. At the linear level the vorticity is
\begin{eqnarray*}
\bm{\omega}(\br)=q s\sin\theta\left(\partial_x\eta_2\vect{e}_y-\partial_y\eta_2\vect{e}_x\right)\nonumber\\
\bm{\omega}(\bk)=iq s\sin\theta\left(k_x\vect{e}_y-k_y\vect{e}_x\right)\eta_2(\bk).
\end{eqnarray*}
Solving Eq.~(\ref{v_fix}) for the velocity gives
\begin{equation*}
\vect{v}(\bk)=\frac{q s\sin\theta}{k^2}\left(k_xk_z\vect{e}_x+k_yk_z\vect{e}_y-k_{\perp}^2\vect{e}_z\right)\eta_2(\bk),
\end{equation*}
so that the linearized advection term is
\[\left(\vect{v}\cdot \nabla\right)\vect{n}=-q^2s\sin^2\theta \frac{k_{\perp}^2}{k^2}\eta_2(\bk)\vect{e}_{1},\]
%
%
leading to the dispersion relation
\begin{equation*}\label{dispersion}
\Omega(k)=q\cos\theta k_z\pm\sqrt{\frac{k^2}{2}\left[\frac{k^2}{2}+q^2\sin^2\theta\left(s\frac{k_\perp^2}{k^2}-\frac{1}{2}\right)\right]}.
\end{equation*}
Note that the growth rate of the unstable modes is now anisotropic, with the transverse modes always stable. This result has been checked in both the spin-1 and spin-1/2 cases by calculating the corresponding Bogoliubov modes of Eq.~(\ref{GP_L}) before taking the incompressible limit $c_0,c_2\to\infty$. An extensive analysis of the Bogoliubov modes of the helix configurations, not restricted to the incompressible limit, was recently given in Ref.~\cite{Cherng2008}.

\section{Dipolar interactions and the Larmor instability}\label{sec:larmor}

We now turn to the question of the stability of Larmor precession. Absent dipolar interactions, rotational symmetry in the spin space means that the only effect of a magnetic field is precession at the Larmor frequency $ \omega_L\equiv g\mu_B H$. We will now show that accounting for dipolar forces renders Larmor precession unstable in general. The approximation of averaging the dipolar interactions over the Larmor trajectories, as was done in Ref.~\cite{kawaguchi2007}, for example, is therefore guaranteed to break down at sufficiently long times.

The simplest geometry to consider is an infinite plane of thickness $d$. If the plane is perpendicular to the $x$-axis, the demagnetizing field for a uniform magnetization is $\vect{M}$ is $-\left(\vect{M}\cdot\vect{e}_x\right)\vect{e}_x$, which is on the order of $10^{-5}$ Gauss for a typical atomic gas. 

For a magnetic field in the $z$-direction, the magnetostatic energy per particle is then $s e_{\mathrm{mag}}$, with
\begin{eqnarray*}
e_{\mathrm{mag}}(\vect{n})=\frac{1}{2}\omega_{\perp} n_x^2+\omega_L n_z,\qquad
\omega_\perp\equiv\mu_0s(g\mu_B)^2 \rho ,
\end{eqnarray*}
where $\rho$ is the density of the gas. The easy-plane anisotropy energy $\omega_\perp$ due to the dipolar forces in on the order of $h\times 10\, \mathrm{Hz}$ under typical experimental conditions. We may use the above energy as a \emph{local} energy density as long as the magnetization varies on sufficiently large scales $\gg d$. As we will see in the next section, in 2d the long-ranged dipolar interactions have the non-analytic form $\sim k_{\alpha}k_{\beta}/|k|$, so that the leading deviation from this approximation is $O(kd)$ i.e. \emph{linear} in the spin wave wavevector. Requiring that this part is small compared to the usual quadratic spin wave dispersion $\omega_k\equiv k^2/2$ then gives the two conditions $\omega_\perp d\ll k\ll d^{-1}$. The equations of motion are then
\begin{equation*}
\frac{D\vect{n}}{Dt}+\vect{n}\times \left(\frac{\partial e_{\mathrm{mag}
}}{\partial \vect{n}}-\frac{1}{2}\nabla^2\vect{n}\right)=0
\end{equation*}
Working in the canonical coordinates $n_z$, $\phi$, the precession of a spatially constant magnetization $\vect{n}_0(t)$ obeys the equations 
\begin{eqnarray}\label{LL_ham}
\dot\phi=\omega_L-\omega_\perp n_z\cos^2\phi =\frac{\partial e_{\mathrm{mag}}(\phi,n_z)}{\partial n_z}\nonumber\\
\dot n_z=\frac{\omega_\perp}{2}\left(1-n_z^2\right)\sin 2\phi =-\frac{\partial e_{\mathrm{mag}}(\phi,n_z)}{\partial \phi}
\end{eqnarray}
which may be solved exactly in terms of elliptic functions. For the case of small angle precession about the direction of the magnetic field, Kittel's classical result for the precession frequency is $\sqrt{\omega_L(\omega_L+\omega_\perp)}$~\cite{kittel1948}. 

Introducing as before the Frenet-Serret frame we find the linearized equation of motion for the spin waves (since the zeroth order solution is constant in space the advection term plays  no role)~\cite{kashuba2006}
%
\begin{equation*}\label{kashuba_sw}
\epsilon_{ab}\partial_t\eta_b=\frac{1}{2}\nabla^2\eta_a+\omega_\perp\left(\eta_a\cos^2\phi(t)-\vect{e}_a^x(t)\vect{e}_b^x(t)\eta_b\right)
\end{equation*}
%
Where we have used the expression for the torsion $\tau(t)=n_z\dot\phi$, and the equation of motion Eq.~(\ref{LL_ham}). Introducing the complex notation $Z_k^{\dagger}=(z^*_k,z_k)$, with $z_k=(\eta_{1k}+i\eta_{2k})e^{-i\Theta_k(t)}$ we find
%
%
%
%
%
\begin{eqnarray}\label{complex_sw}
-i\left(\begin{array}{cc}1 & 0 \\0 & -1\end{array}\right)\partial_tZ_k=\left(\begin{array}{cc}0 &f_k(t) \\f_k^*(t) & 0\end{array}\right)Z_k,\nonumber\\
f_k(t)\equiv-\frac{\omega_{\perp}}{2}e^{-2i\Theta_k(t)}\left[\vect{e}^x_1\vect{e}^x_1-\vect{e}^x_2\vect{e}^x_2+2i\vect{e}^x_1\vect{e}^x_2\right]\nonumber\\
\dot\Theta_k(t)\equiv-\frac{k^2}{2}-\frac{\omega_{\perp}}{2}\left[\vect{e}^x_1\vect{e}^x_1+\vect{e}^x_2\vect{e}^x_2\right]+\omega_\perp\cos^2\phi.
\end{eqnarray}
We illustrate the instability at lowest order in the dipolar interactions. At this order we evaluate the driving term in Eq.~(\ref{complex_sw}) $f_k(t)$ on the unperturbed Larmor precession
%
\[f_k(t)\approx\frac{\omega_\perp}{8}e^{2i\omega_kt}\left[\left(1+n_z\right)e^{i\omega_Lt}-\left(1-n_z\right)e^{-i\omega_L t}\right]^2\]
Evidently there are resonances at $\omega_L=\pm\omega_k$. The size of the resonance region can be obtained with the trial solution $z=\alpha e^{i\omega t}+\beta e^{i\left(2\omega_k\mp2\omega_L-\omega\right)}$. Solving the eigenvalue equation for $\omega$ for the $\omega_k\approx\omega_L$ resonance gives
\begin{equation*}\label{first_res}
\omega=\omega_k- \omega_L\pm\sqrt{\left(\omega_k-\omega_L\right)^2-\frac{\omega_\perp^2}{64}\left(1- n_z\right)^4}.
\end{equation*}
Finite $\omega_\perp$ opens regions of complex $\omega$ in the vicinity of the resonances where spin wave amplitudes grow exponentially. Generally we expect resonances at $\omega_k=n\omega_L$ for integer $n$. The resonances are present for arbitrarily weak dipolar forces, and could provide a novel signature of these interactions even in the limit $\omega_\perp\ll \omega_L$. The long wavelength assumption used in the derivation is likely not necessary for the existence of an instability as both Eq.~(\ref{LL_ham}) for the uniform domain and the spin waves are common to the general case. The instability is a parametric resonance that results from their coupling. The comparatively large Zeeman fields used in current experiments do lead to the difficulty that the thermal spin wave amplitudes at wave vectors corresponding to energies $\sim \omega_L$ that provide the initial `seed' for the instability are very small. At present this is the main obstacle to the observation of the predicted effect.


\section{Spin-wave instabilities with Larmor averaged Dipolar interactions}\label{sec:ave}

\begin{figure}

$k_zd$
\centering \includegraphics[width=0.4\textwidth]{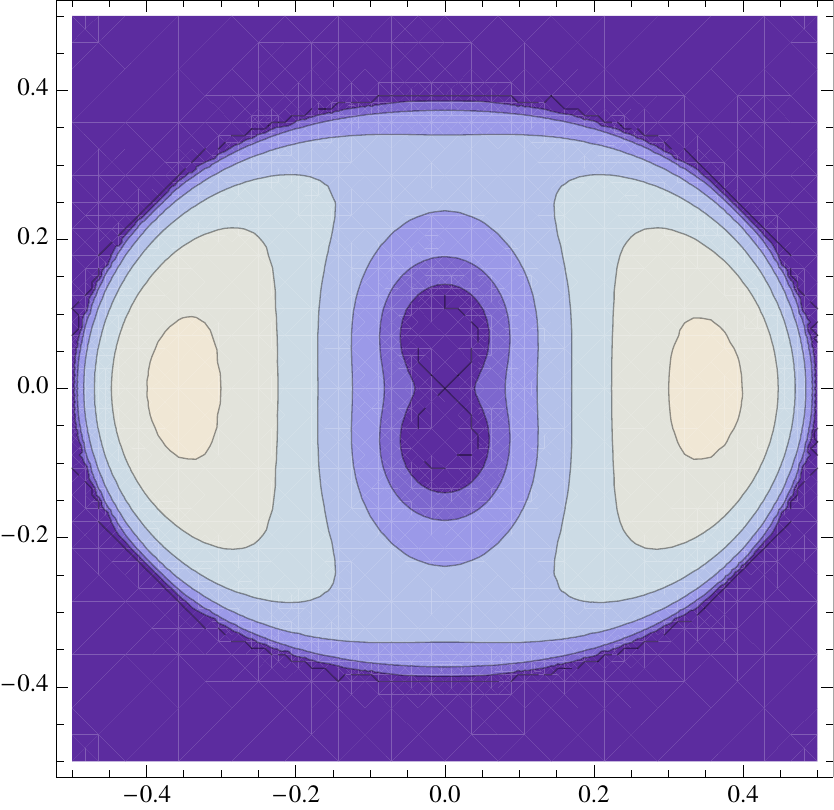} 
\centering \includegraphics[width=0.4\textwidth]{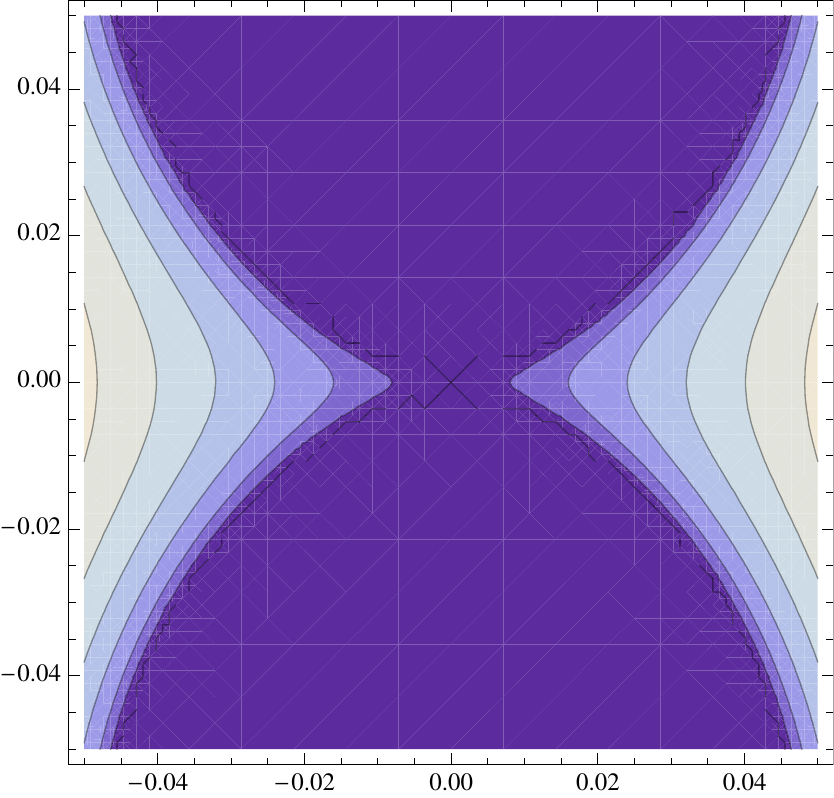}

\centering $k_yd$
\caption{Unstable mode spectrum Eq.~(\ref{lowq}) with $\omega_{\perp}d^2m/\hbar^2=0.25$. The lower figure is a magnification of the region near the origin.
\label{fig:batsignal}}
\end{figure}

If the parametric instabilities discussed in the previous section are unlikely to be seen in present experiments, it is natural to investigate the effect of the Larmor averaged dipole interaction. For a magnetic field in the $z$-direction this takes the form~\cite{kawaguchi2007}
\begin{widetext}
\begin{equation}\label{dd_ave}
H_{\mathrm{dip}}=\frac{g_d}{8\pi}\int d^3\br_1 d^3\br_2\frac{x_{12}^2+y_{12}^2-2z_{12}^2}{ r_{12}^5}\left[n^z(\br_1)n^z(\br_2)-\frac{1}{2}\left(n^x(\br_1)n^x(\br_2)+n^y(\br_1)n^y(\br_2)\right)\right].
\end{equation}
\end{widetext}
where $\br_{12}=\br_1-\br_2$ and $g_d=\mu_0\left(gs\mu_B\right)^2\rho^2$. In this section we will show that the effect of this interaction is to generate a spin wave instability of a system with uniform transverse magnetization. The resulting spectrum of unstable modes has a characteristic anisotropic structure in wavevector space (see Fig.~\ref{fig:batsignal}) similar to that observed in the recent experiment of Ref.~\cite{Vengalattore2008}. This instability has very recently been discussed from the Bogoliubov viewpoint in Ref.~\cite{Cherng2008a}, while we will focus on the long wavelength limit. This approach is not expected to be quantitatively valid everywhere in the unstable region of wavevector space, but it captures the anisotropic structure near zero momentum exactly.

Again we consider an infinite plane of thickness $d$ in the $y-z$ plane. To obtain the momentum space form of the dipolar interactions, it is convenient to make use of the formula
\begin{equation}\label{form}
\frac{x_{12}^2+y_{12}^2-2z_{12}^2}{ r_{12}^5}=\partial_{z_1}\partial_{z_2}\frac{1}{r_{12}}-\frac{4\pi}{3}\delta^{(3)}(\br_{12}),
\end{equation}
with which we obtain
\begin{eqnarray}
H_{\mathrm{dip}}
&=&\frac{g_d d}{8\pi} \int \frac{d^2 q}{\left(2\pi\right)^2}\left(n^z_{\bq}n^z_{-\bq}-\frac{1}{2}\left[n^x_{\bq}n^x_{-\bq}+n^y_{\bq}n^y_{-\bq}\right]\right)\nonumber\\
&&\times\left[q_z^2d^2g(|\bq|d)-\frac{4\pi}{3}\right]
\end{eqnarray}
The second term in the square brackets arises from the $\delta$-function term of Eq.~(\ref{form}) and accounts for the anisotropy energy introduced in the previous section: after Larmor averaging an easy-plane anisotropy an easy axis in the $z$-direction is obtained.  The first term accounts for the finite wavevector correction to this picture and involves the function $g(x)$ with asymptotes
\begin{equation}\label{g_asy}
g(x)\to\begin{cases}
\frac{2\pi}{x} & x\ll 1 \cr
\frac{4\pi }{x^2} & x\gg 1.
\end{cases}
\end{equation}
The low wavevector asymptote displays the non-analyticity characteristic of dipolar forces. 

From the point of view of energetics the easy axis term is the source of the instability that we will discuss: a system that starts with transverse magnetization can lower its dipolar energy by ordering at long wavelengths in the $\pm z$ directions. Incidentally, the asymptote Eq.~(\ref{g_asy}) also shows that a tightly wound helical configuration with magnetization in the $x-y$ plane and axis in the $z$-direction has the same dipolar energy as a uniform configuration aligned in the $z$-direction, but of course the former has a lower kinetic energy.

It is now straightforward to linearize the LLE about uniform transversely magnetized state (again we can ignore the advective term). Using the same notations as in Section~\ref{sec:helix} and considering only the case $\theta=0$ of purely transverse magnetization, we obtain the coupled equations
\begin{eqnarray}\label{LLE_lin_dip}
\dot\eta_{2}(\bk)&=&\left[-\frac{1}{2}k^2 +\frac{\omega_\perp d^2}{8\pi}k_z^2g(kd)\right]\eta_{1}(\bk)\nonumber\\
\dot\eta_{1}(\bk)&=&\left[\frac{1}{2}k^2-\frac{\omega_\perp}{2}+ \frac{\omega_\perp d^2}{4\pi}k_z^2g(kd)\right]\eta_{2}(\bk),
\end{eqnarray}
which give for the unstable modes
\begin{equation}\label{lowq}
\Omega(k)=\sqrt{\left(\frac{k^2}{2}-\frac{\omega_\perp d}{4}\frac{k_z^2}{k}\right)\left(\frac{k^2}{2}-\frac{\omega_\perp}{2}+\frac{\omega_\perp d}{2}\frac{k_z^2}{k}\right)},
\end{equation}
where we have used the low wavevector asymptote of $g(x)$. The imaginary part of the result Eq.~(\ref{lowq}) is plotted in Fig.~\ref{fig:batsignal}. The dimensionless dipolar energy $\omega_{\perp}d^2m/\hbar^2=0.25$ is appropriate to the experiment Ref.~\cite{Vengalattore2008} with peak density $\rho_0=2.3\times 10^{14}\,\mathrm{cm}^{-3}$ and $d$ taken to be the Thomas-Fermi thickness $1.8 \,\mu\mathrm{m}$. 

The main qualitative feature of the spectrum of unstable modes is the `pinched' structure around zero wavevector, with the boundary between stability and instability having the form $k_z\sim k_y^{3/2}$. It seems plausible that the anisotropic power spectrum of the instability studied in Ref.~\cite{Vengalattore2008} is a reflection of this structure. 

A remaining puzzle of that experiment concerns the observation that the anisotropic spin wave instability depends upon the creation of a helical twist of the transverse magnetization by a magnetic field gradient: no instability was observed in the case of a uniform system. On the other hand, the instability discussed above is hardly affected by the presence of the helix, not least because the characteristic scale of the the latter is much larger. It is then natural to suppose that the field gradients used to create the helix additionally generate spin wave fluctuations that provide a large initial `seed' from which the instability grows. It would therefore be useful to study the same problem in a uniform system at higher temperatures, where such fluctuations may be present to begin with.

\section{Conclusion}

We have developed a long wavelength description of ferromagnetic condensates, which considerably simplifies the understanding of dynamical phenomena relative to the time-dependent GP theory. The utility of the approach was illustrated with a number of examples that either have already been
 realized in the ultracold laboratory, or may be in the near future.




I would like to thank Mukund Vengalattore and Dan Stamper-Kurn for useful discussions of the experimental situation.

\end{document}